\documentclass[preprint2]{aastex}
\usepackage{amsmath}
\usepackage{graphicx}
\usepackage{bm}
\usepackage[colorlinks,
            linkcolor=red,
            anchorcolor=blue,
            citecolor=blue
            ]{hyperref}

\newcommand{\unit}[1]{\ensuremath{\,\mathrm{#1}}}
\def\kms{$\rm{km~s}^{-1}$}

\begin{document}

\bibliographystyle{yahapj}

\title{Solar Filament Longitudinal Oscillations along a Magnetic Field Tube with Two Dips}
\author{Yu-Hao Zhou\altaffilmark{1,2}, Li-Yue Zhang\altaffilmark{1,2},
Y. Ouyang\altaffilmark{1,2}, P. F. Chen\altaffilmark{1,2} and C. Fang\altaffilmark{1,2}}
\affil{$^1$ School of Astronomy and Space Science, Nanjing University, Nanjing 210023, China; \email{chenpf@nju.edu.cn}}
\affil{$^2$ Key Laboratory of Modern Astronomy \& Astrophysics (Nanjing University), Ministry of Education, Nanjing 210023, China}

\begin{abstract}
The large-amplitude longitudinal oscillations of solar filaments have been 
observed and explored for more than ten years. Previous studies are 
mainly based on the one-dimensional rigid flux tube model with a single 
magnetic dip. However, it is noticed that there might be two 
magnetic dips, and hence two threads, along one magnetic field line. 
Following the previous work, we intend to investigate the kinematics of the 
filament longitudinal oscillations when two threads are magnetically 
connected, which is done by solving one-dimensional radiative hydrodynamic 
equations with the numerical code MPI-AMRVAC. Two different types of 
perturbations are considered, and the difference from previous works resulting 
from the filament thread-thread interaction is investigated. We find that even 
with the inclusion of the thread-thread interaction, the oscillation period is 
modified weakly, by at most 20\% compared to the traditional pendulum model 
with one thread. However, the damping timescale is significantly 
affected by the thread-thread interaction. Hence, we should take it into 
account when applying the consistent seismology to the filaments where two 
threads are magnetically connected.

\end{abstract}

\keywords{Sun: filaments, prominences - Sun: oscillations - methods: numerical}

\section{Introduction} \label{sect1}
As one type of solar activity, filaments (or prominences) are a relatively 
long-lived phenomenon \citep{tand95}. Unlike solar flares, which last 
for only tens of minutes, solar filaments, except some dynamic active-region 
filaments, can survive for days or weeks \citep{engv04}. Although some 
of them might be due to continual flows of chromospheric plasma along magnetic 
field lines where magnetic dips are not required \citep{wang99, karp01, zou16, 
zou17}, most filaments are believed to be supported by magnetic dips, whose 
upward Lorentz force compensates for the gravity \citep{kipp57, 
arre12}. During the long lifetime, it is inevitable for the filament to be 
disturbed by a nearby flare \citep{jing03} or incoming shock waves 
\citep{shen14}. As a result, the filament would then oscillate with a large 
amplitude. The shaking of the footpoints of the magnetic field lines 
can induce propagating Alfv\'en waves \citep{erde07}, and hence lead to 
small-amplitude longitudinal oscillations of the filament threads, which can 
account for the counterstreamings of the filaments \citep{chen14}.

Filament oscillations have been observed for decades since they were identified
by the winking appearance \citep{rams66}. Winking filaments are visible only
when the oscillation velocity is large, say, larger than 20 \kms \citep{oliv02}.
These large-amplitude oscillations are generally believed to be transverse
ones. Later, spectral observations with higher resolutions discovered filament
oscillations with small amplitude down to 500 m s$^{-1}$ \citep{zhan91}. In
theory, these oscillations are modeled by different eigen modes \citep{joar92a, 
joar92b}, which provides the foundation for prominence seismology 
\citep{sole10}, and magnetic parameters can be derived accordingly 
\citep{arre12}.

In 2003, a new type of filament oscillations was discovered, which are
manifested as the monolithic oscillations of the threads along the thread
direction \citep{jing03}. Since it is believed that filament threads trace the
local supporting magnetic field lines, this type of oscillation is called
large-amplitude longitudinal oscillations \citep{jing03, vrsn07}, in contrast
to the small-amplitude longitudinal oscillations \citep{bash83}. Several
mechanisms were qualitatively proposed to explain the restoring force for such
oscillations, e.g., the field-aligned component of gravity, pressure gradient, 
and magnetic tension force or pressure force \citep{jing03, vrsn07, li12}. With
the geometry of the magnetic field line taken from observations, \citet{zhan12}
performed 1-dimensional (1D) radiative hydrodynamic simulations of a filament
thread. Their simulations reproduced the oscillation period, though the
resulting damping time is longer than that in observations. Noticing that the
gas pressure gradient is much smaller than the gravity, they concluded that
the field-aligned component of gravity is the restoring force for the
longitudinal oscillations. With a parameter survey, \citet{luna12b} verified
that except for the case with extremely shallow dips, the field-aligned
component of gravity generally overwhelms the gas pressure gradient inside the
coronal loop, suggesting that the gravity is the main restoring force.
Therefore, the large-amplitude longitudinal oscillations are considered as the
pendulum mode \citep{luna12a}. Within such a framework, \citet{zhan13} conducted
a parameter survey with radiative hydrodynamic simulations, by which they
derived empirical formulae to relate the oscillation period and damping timescale
to the geometric parameters of the magnetic dip and the perturbation velocity.

It should be pointed out that all of these numerical simulations are based on the assumption that the filament threads are totally independent, and only one
dip is present along one magnetic field line. However, some of the magnetic
field lines may possess more than one dip. For example, \citet{mack09} studied
the effect of a small magnetic element approaching a solar filament through
nonlinear force-free modeling. It is found that some field lines have double
dips although the twist of the field line is less than one turn. The nonlinear
force-free field extrapolation conducted by \citet{jing10} also shows the
existence of double dips in some field lines. Moreover, double dips are a
natural result for some field lines in the head-to-tail model for the filament
formation \citep{mart01}. In the case of double dips along one magnetic flux
tube, the filament threads on the double dips would interact with each other
when oscillating.

From observational point of view, it has been revealed that multiple  
threads in one filament oscillate in unison \citep{yi91, terr02, lin07}, 
indicative of interactions among individual threads. Such 
an interaction between two adjacent threads is analytically and numerically 
investigated by \citet{diaz05} and \citet{diaz06}. These works focus on the 
transverse mode of the filament oscillations, where different threads interact 
through fast-mode magnetoacoustic waves. Fast-mode waves might also be excited 
in filament longitudinal oscillations if the magnetic field is so weak that 
the heavy filament thread moving along the field line can alter the magnetic 
field line \citep{li12}. However, when the magnetic field is strong, filament 
longitudinal oscillations can excite only the slow-mode waves, which propagate 
along the field line. In this case, when there are two magnetic dips that host 
two threads, the two threads would also interact through slow-mode waves. It 
is of great interest to see how the interaction between the magnetically 
connected two threads affects their oscillation behaviors during a 
longitudinal oscillation.  In this paper, we aim to investigate the 
large-amplitude longitudinal oscillations of double threads along one flux 
tube. The paper is arranged as follows. The numerical setup is described in 
Section \ref{sec2}, and the results are presented in Section \ref{sec3}, which 
are followed by discussions in Section \ref{sec4}.

\section{Numerical Method} \label{sec2}

Following previous research, we perform 1D radiative hydrodynamic simulations
to study the longitudinal oscillations of solar filaments, where the
deformation of the magnetic flux tube and the variation along the flux tube 
are ignored. This seems to be a quite reasonable approximation since the 
2-dimensional simulations performed by \citet{luna16} revealed that there are 
no obvious interactions between threads in different field lines during 
longitudinal oscillations. The 1D radiative hydrodynamic equations used in 
this paper are as follows:

\begin{equation}
    \frac{{\partial \rho }}{{\partial t}} + \frac{\partial }{{\partial s}}(\rho v) = 0,
    \label{eq1}
\end{equation}
\begin{equation}
    \frac{\partial }{{\partial t}}(\rho v) + \frac{\partial }{{\partial s}}(\rho {v^2} + p) = \rho {g_\parallel}(s),
    \label{eq2}
\end{equation}
\begin{equation}
    \frac{{\partial E}}{{\partial t}} + \frac{\partial }{{\partial s}}(Ev + pv) = \rho {g_\parallel}v  - {n_e}{n_H}\Lambda (T) + \frac{\partial }{{\partial s}}(\kappa \frac{{\partial T}}{{\partial s}})+ H(s),
    \label{eq3}
\end{equation}

\noindent
where $s$ is the distance along the rigid flux tube, ${g_\parallel}(s)$ is the
field-aligned component of the gravity, and all the rest symbols in the
equations have their usual meanings. With the contributions from hydrogen and
helium, the mass density $\rho$ is related to the number density by
$\rho=1.4m_pn_H$, and the gas pressure is $p = 2.3n_Hk_BT$. The total energy
$E = \rho {v^2}/2+p/(\gamma-1)$, where $\gamma=5/3$ is the adiabatic index.
The right-hand side of equation (\ref{eq3}) includes optically thin radiative
(the second term), heat conduction (the third term) and an additional
volumetric heating term
(the last term). $\Lambda (T)$ is the radiative loss coefficient, whose
distribution can be found in Figure 1 of \citet{xia11}. $\kappa  =
{10^{-6}}{T^{5/2}}\unit{erg}\unit{cm}^{-1}\unit{s}^{-1}\unit{K}^{-1}$ is the
Spitzer-type heat conductivity. The steady background heating term $H(s)$,
which is expressed below, is applied to maintain the hot corona:

\begin{equation}
    {H}(s)=
	\begin{cases}
		E_0{\exp{ ( - s/{H_m})}}, & {s < L/2}; \\
		{E_0}\exp [ - (L - s)/{H_m}], & {L/2 \le s < L}.
	\end{cases}
    \label{eq4}
\end{equation}

The heating rate decays exponentially with the distance, with the base
amplitude being $E_0=3\times {10^{-4}}\unit{erg}\unit{cm^{-3}}\unit{s^{-1}}$
\citep{with77}. The scale length is taken to be ${H_m}=L/2$ \citep{with88}.
This heating is based on the assumption that the energy of the background
heating originates from the photospheric motions \citep{asch02}.

\begin{figure*}
	\centering
	\includegraphics[width=\linewidth]{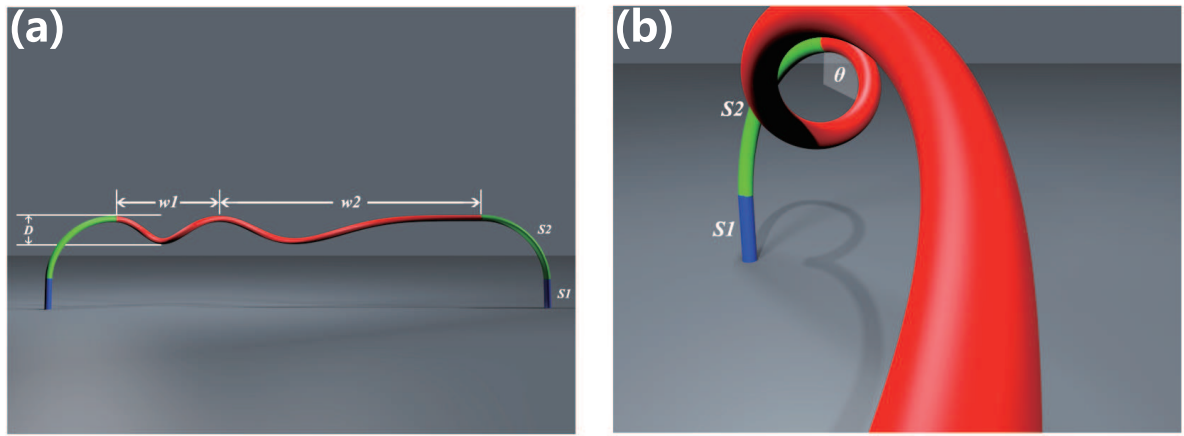}
	\caption{(a) The front view and (b) The side view of the magnetic 
 field configuration used in our work. Different colors represent segments 
 described with different equations. Note that there are two magnetic dips in 
 the displayed magnetic field line, and two separate threads are formed at the 
 two dips.}
    \label{fig1}
\end{figure*}

Note that ${g_\parallel}(s)$ in equation (\ref{eq2}) is determined by the
shape of the flux tube. Following \citet{xia11}, the flux tube is composed of
several parts: A vertical part with a length of $s_1=$ 5 Mm and a
quarter-circle with a length of $s_2$=15.7 Mm at each leg of the field line,
and a central part at the top, as illustrated by Figure \ref{fig1}(a). In the
case of one dip, the shape of the central part is generally described by a
cosine function \citep{xia11, zhan13}. In order to get two dips, we use a helix
with two turns. The curve of the helix is described by the following formulae
in the local coordinates,

\begin{equation}
	\begin{cases}
        x = {(\theta /4\pi )^n}l;\\
        y = 0.5D\sin \theta; \\
        z = 0.5D\cos \theta;
	\end{cases}
    \label{eq5}
\end{equation}

\noindent
where $x$ is the horizontal axis along the main axis of the helix, $y$ is the
horizontal axis perpendicular to $x$, and the $z$-axis is upward, $D$ is the
depth of the dips, $l$ is the axial length of the helix, and $\theta$ ranges
from 0 to $4\pi$. The parameter $n$ controls the shapes of the two dips. When
$n=1$, the two dips are identical. As shown in Figure \ref{fig1}(a), the two
dips have
a length of $w_1$ and $w_2$, respectively, with $w_1+w_2=l$. If $n>1$, then
$w_2$ is larger than $w_1$. The helix is viewed from another angle in Figure
\ref{fig1}(b). In this paper, $D = 3\unit{Mm}$ is chosen, similar to previous
research \citep{anti99, anti00, karp05, karp06, zhan12}. Of course, there can
be many other types of non-uniform helixes with different geometric parameters.
However, as pointed out by \citet{luna12b} and \citet{zhan13}, the oscillation
period is mainly controlled by the curvature radius of the magnetic dip. In this
paper, the curvature radius changes as $l$ changes. Once the geometry of the
flux tube is determined, we can derive ${g_\parallel}(s)$ by $\bm{g}\cdot
\bf{\hat{e}}_s$, where $\bm{g}=-274 \bf{\hat{e}}_z$ m s$^{-2}$,
$\bf{\hat{e}}_z$ is the unit vector of the vertical axis, and
$\bf{\hat{e}}_s$ is the unit vector along the flux tube.

Three steps are taken in order to get the initial conditions with one thread
at each dip. First, we have a hydrostatic loop which is in an equilibrium state.
Similar to \citet{xia11} and \citet{zhan13}, the temperature ($T$) is a
function of height ($z$) with the form

\begin{equation}
    T(z) = {T_1} + \frac{1}{2}({T_2} - {T_1})(\tanh\left( {\frac{{z - {h_0}}}{{{w_0}}}} \right) + 1),
     \label{eq6}
\end{equation}
\noindent
where $T_1=6000\unit{K}$ is the temperature at the chromosphere, whereas
$T_2 = 1\unit{MK}$ is the temperature of the corona, $h_0=2.72$ Mm, and
$w_0=0.25$ Mm. The density distribution is derived from the hydrostatic
equilibrium equation \citep[see][for details]{xia11}. The background heating
can also be derived, which balances the radiation. Second, we need to form two
threads at the two dips. For that purpose, one way is to introduce symmetric
heating localized in the chromosphere, which triggers chromospheric
evaporation, and the ensuing condensation would naturally form threads at the
dips, as demonstrated by \citet{zhan12, zhan13}. For simplicity, however, we
form the threads in a straightforward way, i.e., by increasing the density by
2 orders of magnitude and decreasing the temperature by the same orders along
two segments at the two dips. Each segment is chosen to be $20\unit{Mm}$,
which is the typical length for a filament thread \citep{engv04}. Note that the
filament threads we get in this way are neither in hydrostatic equilibrium nor
thermodynamic equilibrium. As the final step, we start the simulation to let
the non-equilibrium state evolve and relax to an equilibrium state. This same
method has been used in \citet{terr15}, and it is found to perform well.

Once the initial conditions with two segments of threads are determined, the
threads are disturbed to oscillate. There are two types of
perturbations, i.e., a momentum pulse or impulsive heating at one footpoint of
the flux tube. According to \citet{zhan13}, the oscillation properties are not
sensitive to the perturbation type. Therefore, we choose to impose a velocity
perturbation on the filament threads. As mentioned in \S\ref{sect1}, 
the perturbation might be due to a subflare near one footpoint of the magnetic 
field line \citep{jing03} or due to a global coronal wave \citep{shen14}. In 
the former case, only one thread is perturbed initially; In the latter case, 
both threads would be perturbed quasi-simultaneously. In terms of the 
intrinsic magnetic geometry, the two dips can be similar or very different. 
Therefore, in this paper, we consider three cases: In case A, the two dips are 
identical, and the initial perturbation is imposed on one thread; In 
case B, the two dips are different, so the two threads have different 
intrinsic oscillation periods, and the initial perturbation is imposed on one 
thread; In case C, the two threads with different intrinsic oscillation 
periods are perturbed simultaneously. Another case with two identical dips 
and simultaneous perturbations can be ignored since the situation is 
equivalent to two separate threads.

The 1D radiative hydrodynamic equations are numerically solved with the
MPI-Adaptive Mesh Refinement-Versatile Advection Code
\citep[MPI-AMRVAC,][]{kepp12, port14}. In this work, we use 1200 initial grids
with 7 levels of adaptive mesh refinement (AMR), which leads to an effective
resolution ranging from $1.6$ to $4.2\unit{km}$ in different cases. During the
simulations, all the quantities are fixed at the boundaries 
(corresponding to the photosphere) since the dynamics
in the corona has little influence on the two footpoints of the flux tube, as
revealed by \citet{xia11}, \citet{luna12a}, and \citet{zhan13}.

\section{Numerical Results} \label{sec3}

\subsection{Case A: Active and passive threads on two identical dips}

This is an ideal case where the two magnetic dips are identical, i.e., $w_1=
w_2=l/2$, therefore, the intrinsic oscillation periods of the two threads are
the same. The initial velocity perturbation is imposed on the first thread,
which is called ``active thread" hereafter. The velocity perturbation has the
following form:

\begin{equation}
    v(s)=
	\begin{cases}
        0, & s < {s_{1l}} - \delta; \\
        {v_0}(s - {s_{1l}} + \delta )/\delta, & {s_{1l}} - \delta \le s \le {s_{1l}}; \\
        {v_0}, & {s_{1l}} \le s \le {s_{1r}}; \\
        {v_0}( - s + {s_{1r}} + \delta )/\delta, & {s_{1r}} \le s \le {s_{1r}} + \delta; \\
        0, & s > {s_{1r}} + \delta;
	\end{cases}
    \label{eq7}
\end{equation}

\noindent
where ${s_{1l}}$  and ${s_{1r}}$ are the left and right boundaries of the 
first filament thread, respectively; $\delta$ is the thickness of the 
transition layer which allows the velocity perturbation to vary smoothly along 
the flux tube. We set $w_1=w_2=l/2=80\unit{Mm}$, which are also comparable 
with previous works \citep{xia11, zhan13, zhou14}. It is noted that, 
as demonstrated by \citet{zhan13}, the oscillation behavior is not sensitive 
to the form of perturbation. So the velocity profile can be chosen to be 
different from equation (\ref{eq7}), and the results would not be changed 
significantly. 

Figure \ref{fig2} displays the evolution of the temperature distribution
along the flux tube, where the blue segments centered at $s=63$ Mm and $s=141$
Mm correspond to the two threads. It is seen that the first thread at $s=63$
Mm (or active thread) starts to oscillate from the very beginning. As it
moves, sound waves are excited, which propagate with a velocity of
$116\unit{km}\unit{s}^{-1}$ along the flux tube, as indicated by the arrows.
When the first sound wave reaches the second thread (or passive thread) at
$s=141$ Mm, the passive thread starts to oscillate as well, but with a much
smaller amplitude.

In order to study the oscillations more quantitatively, we trace the centers of
the two threads at each time step. The evolutions of the thread centers
are plotted in Figure \ref{fig3} as dashed lines, where the top panel
corresponds to the active oscillation of the first thread and the bottom panel
corresponds to the passive oscillation of the second thread. It reveals that
the active oscillation decays rapidly, whereas the passive oscillation is
initially strengthened, gaining energy from the active thread, and then decays
slowly. Similar to previous authors \citep{jing03, vrsn07, zhan12}, we fit the
oscillations of the two threads with decayed sine functions:

\begin{equation}
    \Delta s_1=A_1{\rm e}^{-t/\tau _1}\sin (\frac{{2\pi}}{p_1}t+\phi_1),
    \label{eq8}
\end{equation}
\begin{equation}
    \Delta s_2=A_2{\rm e}^{-t/\tau_2}\sin (\frac{{2\pi}}{p_2}t+\phi_2),
    \label{eq9}
\end{equation}

\noindent
where $A$, $p$, and $\phi$ are the amplitude, period, and phase of the
oscillation, and the subscripts 1 and 2 refer to the oscillations of the first
(active) and the second (passive) thread, respectively. Note that for the
passive thread, its oscillation begins to damp after the second peak, 
i.e., since $t=86$ min, hence
the corresponding fitting starts at $t=86$ min. The fitted curves are
overplotted on Figure \ref{fig3} as blue solid lines. It is seen that the
oscillations of both threads are fitted very well by the decayed sine
functions, with $A_1$=16.4 Mm, $p_1$=60 min, $\tau_1$=89 min, and
$A_2$=3.4 Mm, $p_2$=58 min, $\tau_2$=230 min.

In order to check how these oscillation properties change with the geometry of
the flux tube, we perform six other simulations with $l/2$ being 50 Mm, 60 Mm,
70 Mm, 90 Mm, 100 Mm, and 110 Mm. Such a range is chosen on purpose: If $l/2$
is too small, there would be direct exchange of materials between the two
threads, whereas if $l/2$  is too large, the dips are too shallow so that the
oscillations would deviate from the pendulum model controlled by gravity
\citep{luna12a}. The corresponding results are presented in Figure \ref{fig4},
where the red triangles represent the active thread, and the blue triangles
represent the passive thread. Though not shown here, similar to
$l/2=80\unit{Mm}$, the amplitudes of the active oscillations in these cases
are 4--5 times larger than those of the passive oscillations.
Whereas the oscillation periods are very close to each other, the damping
timescale of the passive oscillation is generally 2--2.5 times longer than that
of the active oscillation. In order to understand the effect of thread-thread
interactions, we further perform simulations of a single-dipped flux tube,
with the geometry of the magnetic dip, including the depth of the length, being
the same as that of the active thread or passive thread. Note that since the
damping timescale is also sensitive to the oscillation amplitude \citep{zhan13},
we keep the same oscillation amplitude for each corresponding simulation. The
oscillation periods and the damping timescales in the single-dipped simulations
are overplotted in Figure \ref{fig4} as green circles (solid circles for the
active thread and open circles for the passive thread). It is seen that the
oscillation periods of both active and passive threads are close to those of
the corresponding single-dipped simulations, and the damping timescales of the
active oscillations are also close to the results of the single-dipped
simulations, though slightly longer. However, the damping timescales of the
passive oscillations are about 2.5--3 times longer than those of the
single-dipped simulations.

\subsection{Case B: Active and passive threads on two different dips}

In observations, the two dips along one flux tube are probably not
identical, with very different curvature radii. To construct a flux tube with
non-identical dips, we take $n$ in equation (\ref{eq5}) to be 1.322, so that
the length ratio of the two dips is $w_2/w_1\approx 1.5$, which leads to the
curvature radius of the second dip being twice that of the first dip. Note that
the total length of the helix, $l$, is adjusted in order to make sure that the
curvature radius of the first dip (with a shorter length) in case B is the same
as that in case A.

In this case, the same type of velocity perturbation is imposed on the first
thread (or the active thread, the one in the shorter dip) as in case A.
Similarly, in response to the sound waves excited by the active thread, the
passive thread begins to oscillate, with an amplitude about 1/5 of the active
oscillation. Similar to case A, both threads show damped oscillations, as
revealed by Figure \ref{fig5}, which shows the evolutions of the
temperature distributions in case B with $w_1=76$ Mm and $w_2=114$ Mm.

Again, we simulate six other cases with different lengths of the magnetic dips,
and fit the oscillation profiles with the decayed sine functions as described
by equations (\ref{eq8}--\ref{eq9}). The resulting periods and the damping
timescales are displayed in Figure \ref{fig6} as functions of the length of
the dips. In this figure, the red triangles in the top row correspond to the
active thread, and the blue triangles in the bottom row correspond to the
passive thread. In order to check the effect of the thread-thread interaction,
we also perform simulations of the single-dipped oscillations for both the
active and passive threads. The corresponding results are represented by solid
and open circles in Figure \ref{fig6}. It is found that for the active thread,
the oscillation period is almost the same as in the single-dipped flux tube,
with the difference becoming evident only when the length of the dip, $w_1$,
is larger than 95 Mm, but the damping timescale is significantly smaller than
that in the single-dipped case. The relative difference increases from 10\% to
33\% when $w_1$ increases from 47 Mm to 95 Mm. For the passive thread which
has a larger length, $w_2$, both the oscillation period and the damping
timescale deviate from the results of the single-dipped flux tube. It is seen
that as the length of the dip, $w_2$, increases, the oscillation period
becomes smaller and smaller than that in the single-dipped case. When
$w_2$=156 Mm, the oscillation period drops by 16\%. The variation of the
damping timescale is remarkable, as indicated by Figure \ref{fig6}(d).
Contrary to the case of a single-dipped flux tube where $\tau_2$ decreases
with $w_2$, the damping timescale of the passive thread in our case B
increases with $w_2$ almost monotonically.

In the above simulations in this subsection, the velocity perturbation is
imposed on the thread inside the shorter dip. It is therefore worth
investigating the case with the initial perturbation imposed on the thread
inside the longer dip. Note that the geometry of the flux tube is kept the
same as in previous simulations in this subsection.

As shown in Figure \ref{fig7}, the passive thread (the thread in the shorter
dip) starts to oscillate as a result of the sound waves excited by the active
thread. The passive oscillation in this case has a slightly lower amplitude
than the passive oscillation in Figure \ref{fig5}, which is about 1/6 of
the amplitude of the active oscillation. Both threads experience a damped
oscillation after the passive oscillation reached its peak at about $t=53$ min.

Once again, we simulate six other cases with different lengths of the magnetic
dips. Figure \ref{fig8} displays the resulting periods and damping timescales
as functions of the lengths of the dips. The symbols used here are the same as
in Figure \ref{fig6}, but the subscript 1, corresponding to the first thread,
represents the passive oscillations, and the subscript 2, corresponding to the
second thread, represents the active thread. The green circles in Figure
\ref{fig8} also represent the results calculated from the single-dipped
oscillations in order to check the effect of the thread-thread interaction.
In contrast to the period of the passive oscillation in Figure \ref{fig6}, it
is found that the oscillation periods of the passive thread in Figure
\ref{fig8}(a) are almost the same as those of the single-dipped case. However,
the oscillation periods of the active thread in Figure \ref{fig8}(c) are
about 1/4 smaller than those of the single-dipped case. The damping
timescale of the the passive threads in Figure \ref{fig8}(b) shows a tendency
similar to the passive thread in Figure \ref{fig6}(d), i.e., $\tau$ increases
with increasing $w$ almost monotonically. However, the damping timescales of
the active thread in Figure \ref{fig8}(d) are 60--70\% longer than those in
the single-dipped cases, which is opposite to the situation in Figure
\ref{fig6}(b).

\subsection{Case C: Non-uniform helix with concurrent perturbations}

In this case, the geometry of the helix is the same as in case B, but the same
velocity perturbation as described by equation (\ref{eq7}) is also imposed on
the second thread simultaneously. Note that $s_{1l}$ and $s_{1r}$ are replaced
with $s_{2l}$ and $s_{2r}$ for the second thread.

The dashed lines in Figure \ref{fig9} describe the oscillations of the two
threads in the case with $w_1$=76 Mm and $w_2$= 114 Mm, where the top panel
corresponds to the thread with a shorter dip and the bottom panel corresponds
to the thread with a longer dip. It is seen that whereas the oscillation of
the second thread (with the length of dip $w_2$) can be well fitted with a
decayed sine function, the oscillation of the first thread (with the length of
dip $w_1$) decays rapidly in the first two periods, and then keeps almost the
same amplitude like a decayless oscillation. The profile can be fitted neither
with a decayed sine function nor a decayed Bessel function which was used to
describe large-amplitude longitudinal oscillations by \citet{luna12b}.
Therefore, we use the decayed sine functions to fit all the simulation data
for the second thread and the early data (i.e., in the first two periods) for
the first thread. The fitted functions are overplotted in Figure \ref{fig9} as
blue curves.

Similar to cases A and B, we perform other simulations with different lengths 
of the magnetic dips. Figure \ref{fig10} displays how the fitted periods and 
damping timescales change with the length of the dips. We can see that 
for the first thread (top row), the oscillation periods (red triangles) are 
14--20\% higher than those in the single-dipped flux tube (green solid 
circles). Its initial damping timescales during the first two periods (red 
triangles) are remarkably smaller than those in the single-dipped flux tube 
(green solid circles). For the second thread (bottom row), the oscillation
periods are roughly the same as in the single-dipped flux tube (only when
$w_2$ becomes large, the discrepancy becomes significant). However, the 
damping timescales (blue triangles) are significantly larger than those in the 
single-dipped situation (green open circles).

\section{Discussions} \label{sec4}

Waves or oscillations carry important information about the local magnetic
field and/or plasma, and both the oscillation period and the damping timescale
can be used to diagnose the thermal and magnetic structures in the corona
\citep{andr09}. The same as other types of waves or oscillations, the
large-amplitude longitudinal oscillations of solar filaments have also been
applied to derive the magnetic properties of the filaments \citep{luna12b,
zhan13}. However, all the previous simulation work was based on the assumption
that there is only one filament thread along each magnetic field line. Either
from theoretical or extrapolation points of view, some magnetic field lines
may have two or more dips, supporting two or more threads along one flux tube.
In this case, the oscillation of each thread is not independent, and the
interactions between the threads are expected to change the oscillation
behaviors of each thread. With 1D radiative hydrodynamic simulations of the
oscillations of two threads along one flux tube, we showed that indeed the
oscillation properties are changed a lot, compared to the case of a
single-dipped flux tube.

\begin{figure}
	\centering
	\includegraphics[width=\linewidth]{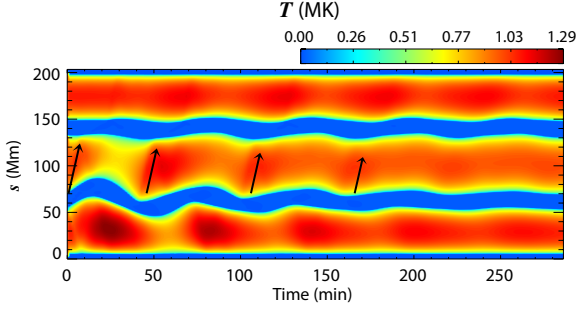}
	\caption{Evolution of the temperature distribution along the magnetic flux tube in case A after the velocity perturbation is imposed on the thread centred at 63 Mm. Black arrows represents energy transfer done by sound waves.}
    \label{fig2}
\end{figure}

\begin{figure}
	\centering
	\includegraphics[width=\linewidth]{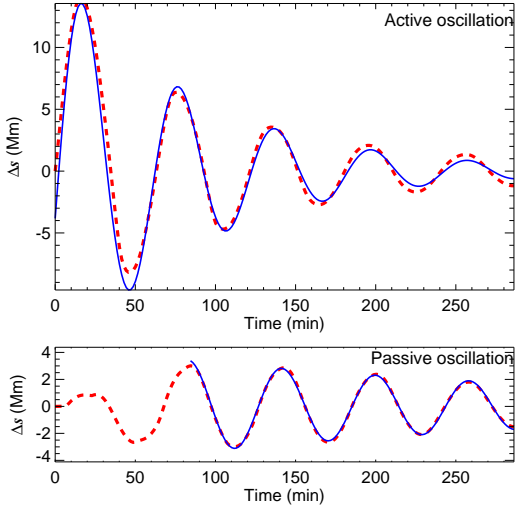}
	\caption{Oscillations of the two filament threads in case A, with the 
 upper panel for the first (active) thread and the lower panel for the second 
 (passive) thread, respectively. The red dashed curves are the spatial 
 displacement of the mass centres of the two filament threads in our numerical 
 simulations, and the blue solid lines are the fitting curves generated by the 
 decayed sine functions.}
    \label{fig3}
\end{figure}

\begin{figure}
	\centering
	\includegraphics[width=\linewidth]{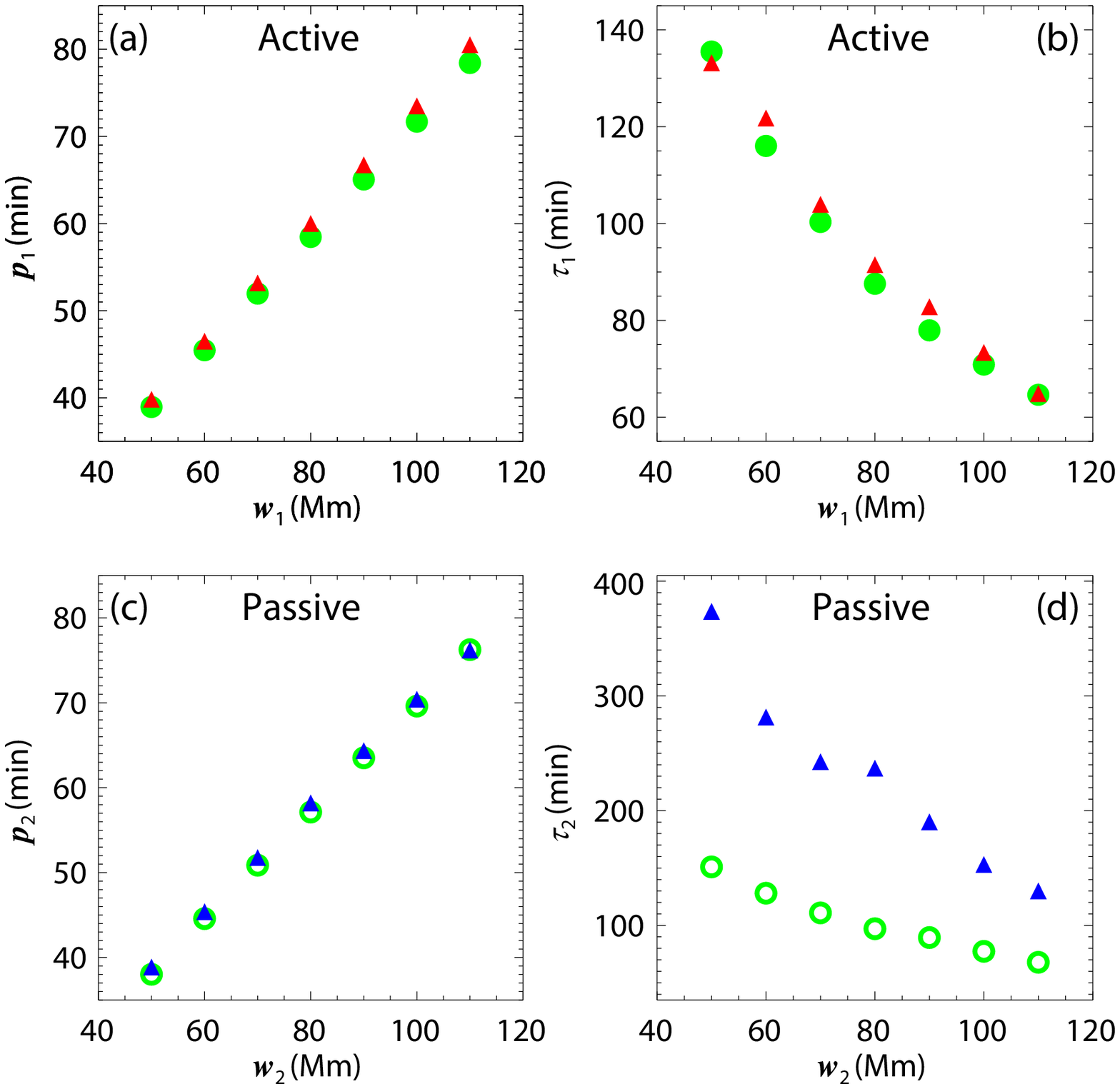}
        \caption{Scatter plots of the period (panel a) and the damping 
 timescale (panel b) of the active thread ({\it red triangles}) in case A; 
 Scatter plots of the period (panel c) and the damping timescale (panel d) of 
 the passive thread ({\it blue triangles}) in case A. Green (solid and open) 
 circles are the results from the corresponding single-dipped simulations.}
    \label{fig4}
\end{figure}

\begin{figure}
	\centering
	\includegraphics[width=\linewidth]{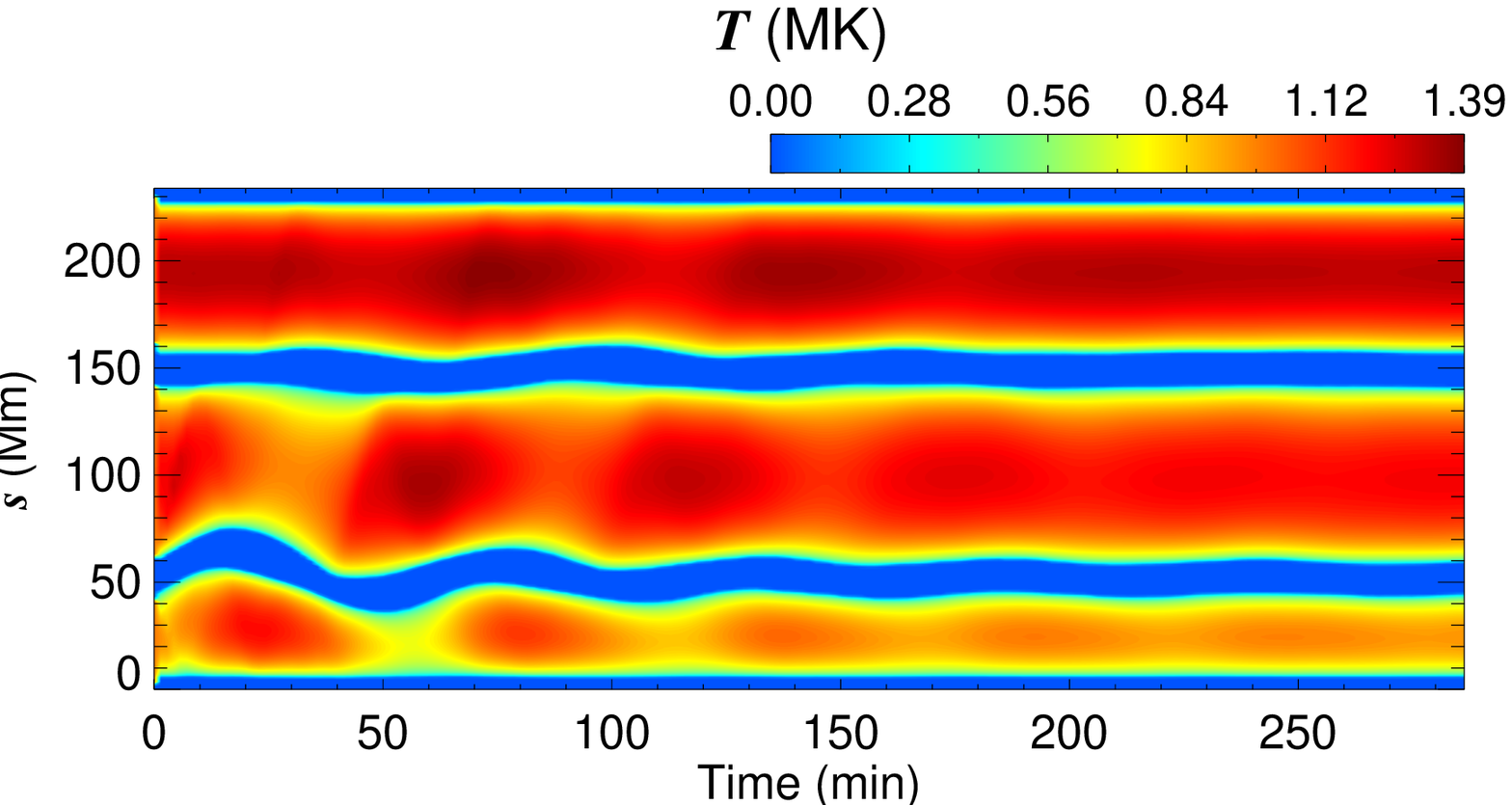}
	\caption{Evolution of the temperature distribution along the magnetic 
 flux tube in case B after the velocity perturbation is imposed on the thread 
 centred at $s=$52 Mm.}
    \label{fig5}
\end{figure}

\begin{figure}
	\centering
	\includegraphics[width=\linewidth]{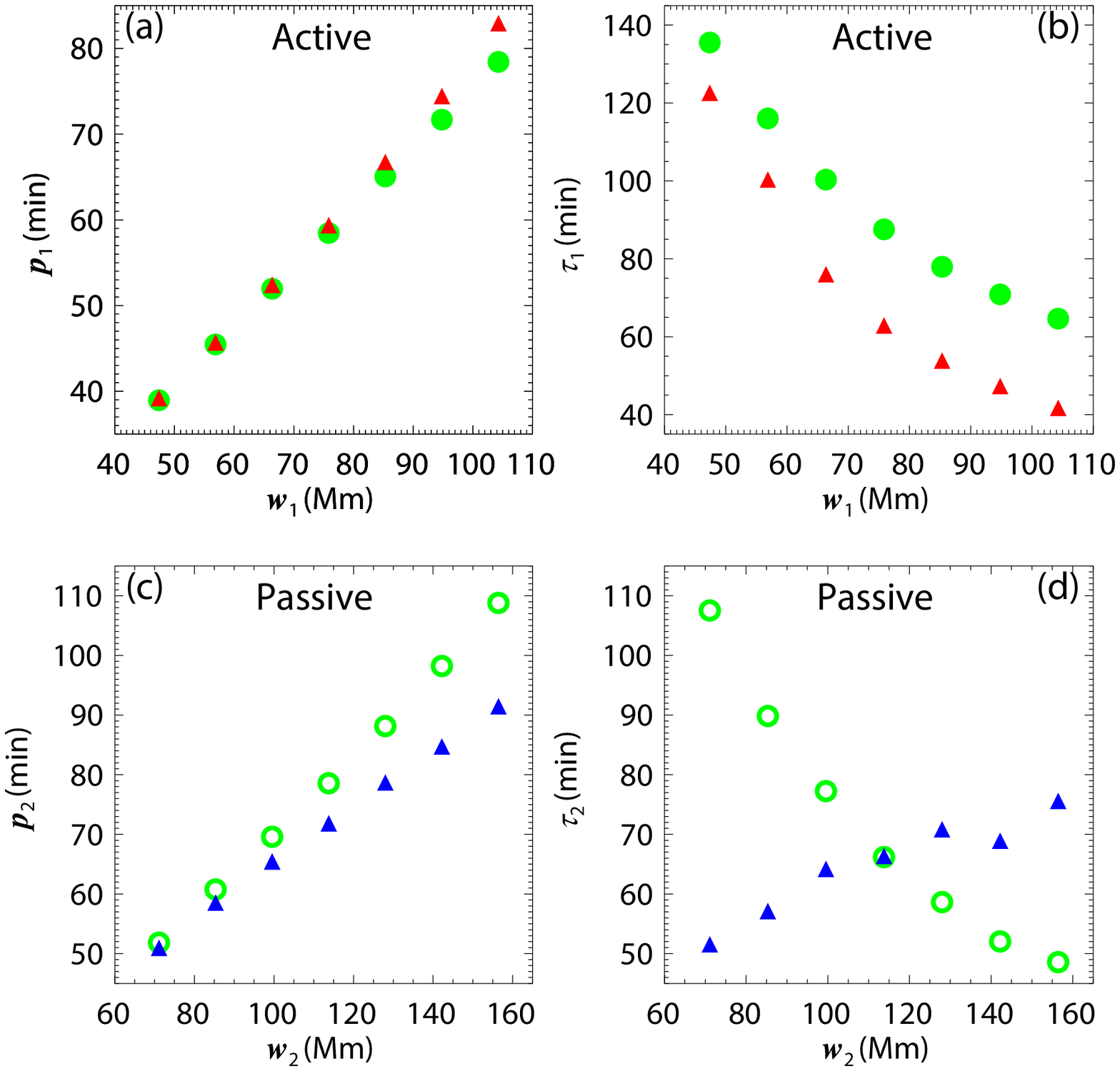}
	\caption{Scatter plots of the period (panel a) and the damping 
 timescale (panel b) of the active thread ({\it red triangles}) in case B; 
 Scatter plots of the period (panel c) and the damping timescale (panel d) of 
 the passive thread ({\it blue triangles}) in case B. Green (solid and open) 
 circles are the results from the corresponding single-dipped simulations.}
    \label{fig6}
\end{figure}

\begin{figure}
	\centering
	\includegraphics[width=\linewidth]{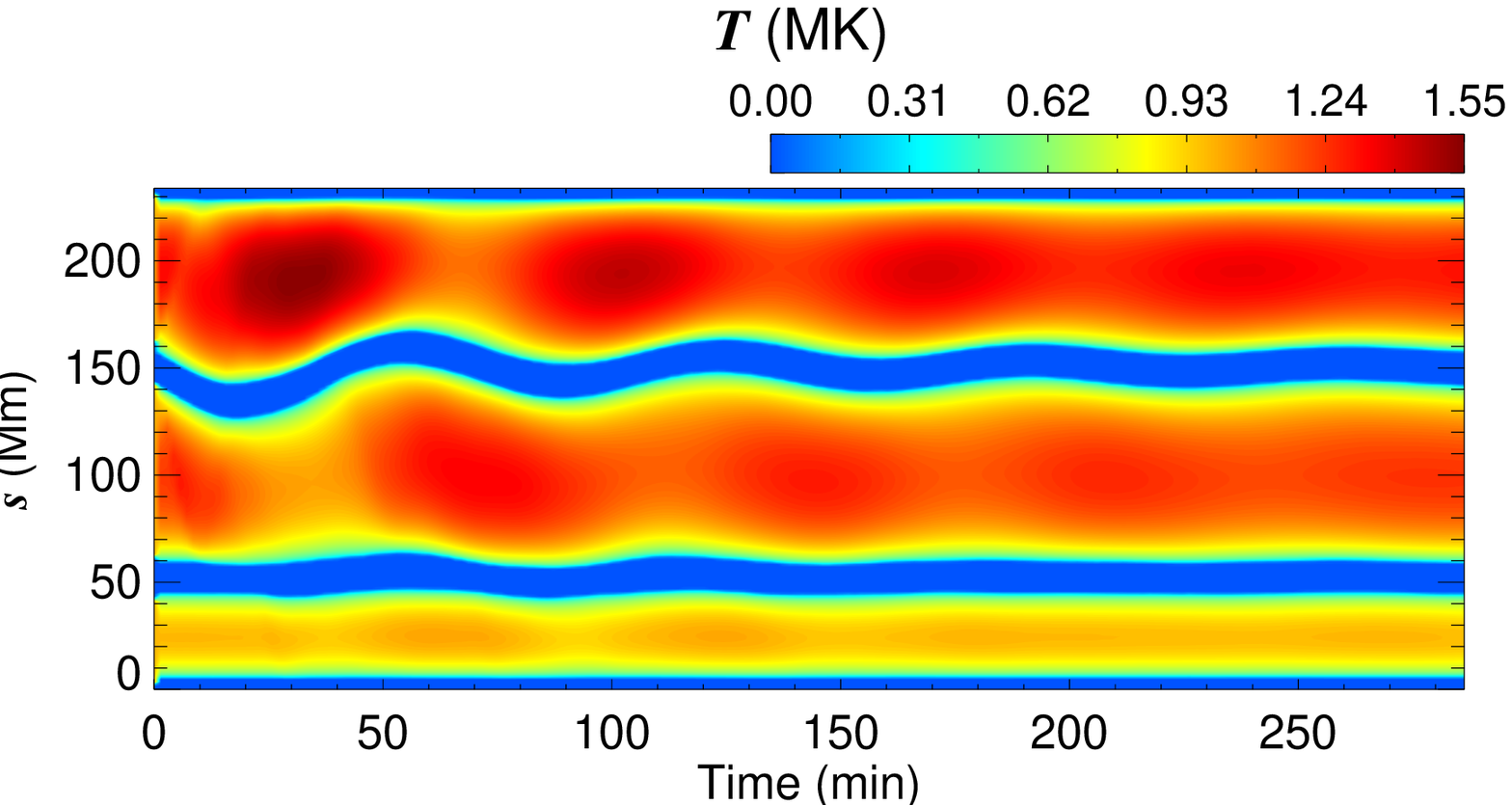}
	\caption{Evolution of the temperature distribution along the magnetic 
 flux tube in case B after the velocity perturbation is imposed on the thread 
 centred at $s=$152 Mm.}
    \label{fig7}
\end{figure}

\begin{figure}
	\centering
	\includegraphics[width=\linewidth]{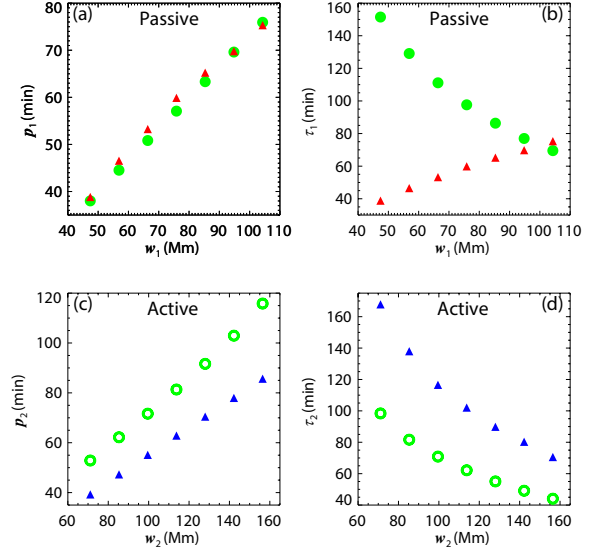}
	\caption{The same as Fig. \ref{fig6}, but the active thread where the 
 initial perturbation is imposed is the one at the longer magnetic dip.}
    \label{fig8}
\end{figure}

\begin{figure}
	\centering
	\includegraphics[width=\linewidth]{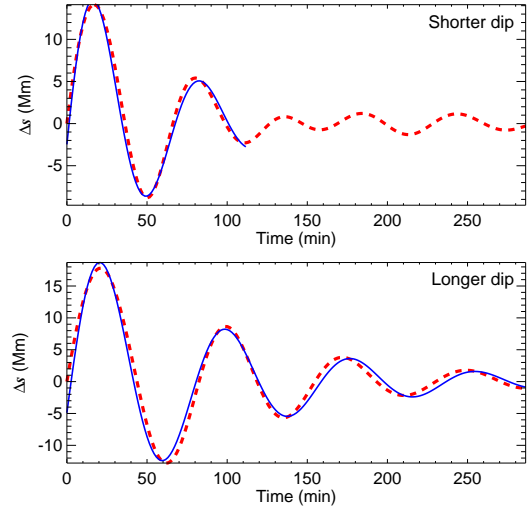}
	\caption{Oscillations of the two filament threads in case C, with the 
 upper panel for the thread at the shorter magnetic dip and the lower panel 
 for the thread at the longer dip, respectively. The red dashed curves are the 
 spatial displacement of the mass centres of the two filament threads in our 
 numerical simulations, and the blue solid lines are the fitting curves 
 generated by the decayed sine functions.}
    \label{fig9}
\end{figure}

\begin{figure}
	\centering
	\includegraphics[width=\linewidth]{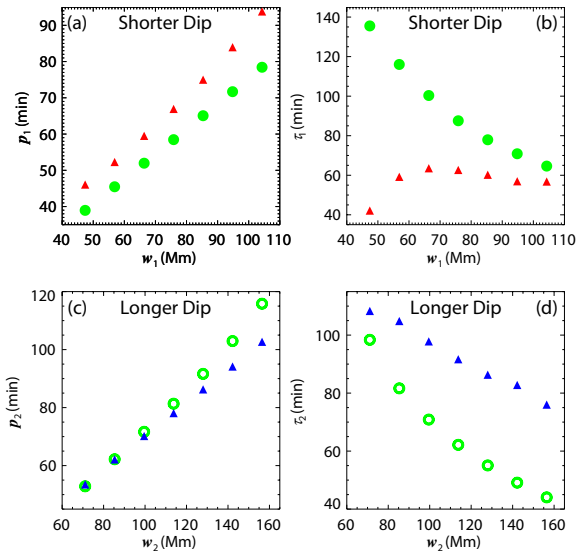}
	\caption{Scatter plots of the period (panel a) and the damping 
 timescale (panel b) of the thread at the shorter magnetic dip ({\it red 
 triangles}) in case C; Scatter plots of the period (panel c) and the damping 
 timescale (panel d) of the thread at the longer dip ({\it blue triangles}) in 
 case C. Green (solid and open) circles are the results from the corresponding 
 single-dipped simulations.}
    \label{fig10}
\end{figure}

For the special case where the two dips are identical and one thread is
perturbed initially (i.e., case A), energy is transferred from the active 
thread to
the passive thread, as indicated by the temperature front which is marked by
the arrows in Figure \ref{fig2}. As a result, the oscillation of the passive
thread decays very slowly, with the damping timescale 2--2.5  times longer
than in the case of a single-dipped flux tube. The decay of the oscillation of
the active thread is not much affected. This is understandable since only 4\%
of its energy is transferred to the passive thread, as implied by the ratio of
the amplitudes of the active and passive threads being about 5. It is also
noticed that although the oscillation of the passive thread is delayed
initially with respect to the active thread due to limited propagation
velocity of sound waves, the oscillation of the two threads tend to be
synchronous as implied by the fifth peaks of both oscillations in Figure
\ref{fig3}. This means that the interaction between the two threads 
leads to the synchronization of their oscillations.

In the case where the curvature radius of the passive thread is twice that of
the active thread, it seems that the oscillation period of the thread inside
the shorter dip is barely affected by the thread-thread interaction, no matter
it is active or passive oscillation. However, for the thread inside the longer
dip, its oscillation period is significantly reduced, i.e., by 26\% when
compared to the single-dipped case, especially when it is the active thread.
Only when its length is below 100 Mm and it is the passive thread, the 
oscillation period is not affected by the thread-thread interaction.
This is probably because the gas pressure gradient overtakes the field-aligned
gravity when the length of the dip is large, as demonstrated by
\citet{luna12a}. The short-period oscillation of the shorter thread continues
to excite sound waves, which serve as the external driving force for the
longer thread, making the oscillation period of the latter closer to the
period of the shorter thread. The modification of the damping timescale is
remarkable. Its variation with the length of the magnetic dip depends strongly
on whether it is the active or passive thread. For the active thread, its
damping timescale ($\tau$) decreases with the increasing length of the
magnetic dip, the same as in the single-dipped case, though $\tau$ may be
reduced or increased in comparison to the single-dipped case, depending on
whether the magnetic dip is shorter or longer; For the passive thread, its 
damping timescale ($\tau$) increases with the increasing length of the 
magnetic dip, completely opposite to the single-dipped case.

There are various types of perturbations in the dynamic solar corona, including
the ceaseless convection motions or $p$-mode oscillation near the solar
surface, trapped waves from solar flares \citep{liu12}, global coronal waves
driven by coronal mass ejections \citep{chen09}. If it is due to a nearby
subflare \citep{vrsn07}, one segment of thread is triggered to oscillate,
which would drive another magnetically-connected segment of thread to
oscillate, as discussed above. However, when a global coronal wave sweeps the
filament \citep{liu12, shen14}, both segments would be triggered to oscillate
nearly simultaneously. Our simulations of such a case indicate that the
oscillation period of the thread with a smaller curvature radius (hence
shorter period) becomes 15--20\% larger, whereas the oscillation period of the
thread with a larger curvature radius (hence longer period) is reduced
slightly. The oscillation of the first thread decays rapidly in the first two
periods, and then experiences a decayless stage thereafter. It means that
kinetic energy of the thread with a smaller curvature radius is initially
transferred significantly to the other thread, leading to the strong damping
of itself and the slow damping of the other thread with a larger curvature 
radius. In the later stage, it seems that kinetic energy is transferred back 
to the first thread, leading to a decayless oscillation of the first thread.

It has been proposed that both the oscillation period and the damping 
timescale can be utilized for consistent seismology \citep[see][for a 
review]{arre12}. In terms 
of longitudinal oscillations of solar filaments, \citet{zhan13} have done a
parameter survey and derived empirical formulae to relate the oscillation
period and damping timescale to the geometric parameters of the magnetic dips.
Based on the simulations in this paper, it becomes clear that we have to be
careful when applying the empirical formulae to observations, especially
when one thread is magnetically connected to another thread. The results of
this paper can be summarized as follows: It seems that the period of filament 
longitudinal oscillation is not affected too much by the interaction 
between two threads, except the case of the active thread inside the longer
dips, where the gas pressure gradient becomes unnegligible \citep{luna12a}.
Even in the latter case, the oscillation period is reduced by only 26\% compared to the case of a single-dipped flux tube.
However, the situation of the  damping timescale is complex, which might be modified significantly by the interaction between magnetically-connected threads.
When the two dips are identical (or similar) and only the active thread is perturbed initially, the damping timescale of the active thread is changed significantly, which
decreases by 25\% if the active thread has a shorter magnetic dip and increases by 75\% if the active thread has a longer dip.
More importantly, the thread-thread interaction completely changes the behavior of the damping
timescale of the passive thread, which increases with the increasing length of the magnetic dip, a result opposite to the case of a single-dipped flux tube.
When the two threads on non-identical dips are perturbed simultaneously, the
oscillation of the thread with a shorter period cannot be described by a
decayed sine function nor a decayed Bessel function. It decays drastically
during the first two periods, and then becomes kind of decayless. The thread
with a longer period, however, can be well fitted with a decayed sine
function, with the damping timescale increased owing to the thread-thread
interaction.

It is noted here that the results presented in this paper are based on 
a simple model, with two segments of filament threads situated at two magnetic 
dips along one rigid flux tube. All the conclusions are worthy to be verified 
in future by two- or three-dimensional simulations.

\acknowledgments
ZYH thanks Mr. T. Shi from University of Michigan for his help with scripting 
in Adobe$^{\circledR}$ Photoshop$^{\circledR}$ CC. The numerical calculations 
in this paper were performed on the cluster system in the High Performance 
Computing Center (HPCC) of Nanjing University. This research was supported by 
the Chinese foundations (NSFC 11533005 \& 11025314) and Jiangsu 333 Project.

\bibliography{reference}

\begin{thebibliography}{}
\providecommand\natexlab[1]{#1}
\providecommand\JournalTitle[1]{#1}

\bibitem[{{Andries} {et~al.}(2009){Andries}, {van Doorsselaere}, {Roberts},
  {Verth}, {Verwichte}, \& {Erd{\'e}lyi}}]{andr09}
{Andries}, J., {van Doorsselaere}, T., {Roberts}, B., {et~al.} 2009,
  \href{http://dx.doi.org/10.1007/s11214-009-9561-2}{\JournalTitle{\ssr}, 149,
  3}

\bibitem[{{Antiochos} {et~al.}(2000){Antiochos}, {MacNeice}, \&
  {Spicer}}]{anti00}
{Antiochos}, S.~K., {MacNeice}, P.~J., \& {Spicer}, D.~S. 2000,
  \href{http://dx.doi.org/10.1086/308922}{\JournalTitle{\apj}, 536, 494}

\bibitem[{{Antiochos} {et~al.}(1999){Antiochos}, {MacNeice}, {Spicer}, \&
  {Klimchuk}}]{anti99}
{Antiochos}, S.~K., {MacNeice}, P.~J., {Spicer}, D.~S., \& {Klimchuk}, J.~A.
  1999, \href{http://dx.doi.org/10.1086/306804}{\JournalTitle{\apj}, 512, 985}

\bibitem[{{Arregui} {et~al.}(2012){Arregui}, {Oliver}, \& {Ballester}}]{arre12}
{Arregui}, I., {Oliver}, R., \& {Ballester}, J.~L. 2012,
  \href{http://dx.doi.org/10.12942/lrsp-2012-2}{\JournalTitle{Living Reviews in
  Solar Physics}, 9, 2}

\bibitem[{{Aschwanden} \& {Schrijver}(2002)}]{asch02}
{Aschwanden}, M.~J., \& {Schrijver}, C.~J. 2002,
  \href{http://dx.doi.org/10.1086/341945}{\JournalTitle{\apjs}, 142, 269}

\bibitem[{{Bashkirtsev} {et~al.}(1983){Bashkirtsev}, {Kobanov}, \&
  {Mashnich}}]{bash83}
{Bashkirtsev}, V.~S., {Kobanov}, N.~I., \& {Mashnich}, G.~P. 1983,
  \href{http://dx.doi.org/10.1007/BF00145584}{\JournalTitle{\solphys}, 82, 443}

\bibitem[{{Chen}(2009)}]{chen09}
{Chen}, P.~F. 2009,
  \href{http://dx.doi.org/10.1088/0004-637X/698/2/L112}{\JournalTitle{\apjl},
  698, L112}

\bibitem[{{Chen} {et~al.}(2014){Chen}, {Harra}, \& {Fang}}]{chen14}
{Chen}, P.~F., {Harra}, L.~K., \& {Fang}, C. 2014,
  \href{http://dx.doi.org/10.1088/0004-637X/784/1/50}{\JournalTitle{\apj}, 784,
  50}

\bibitem[{{D{\'{\i}}az} {et~al.}(2005){D{\'{\i}}az}, {Oliver}, \&
  {Ballester}}]{diaz05}
{D{\'{\i}}az}, A.~J., {Oliver}, R., \& {Ballester}, J.~L. 2005,
  \href{http://dx.doi.org/10.1051/0004-6361:20052759}{\JournalTitle{\aap}, 440,
  1167}

\bibitem[{{D{\'{\i}}az} \& {Roberts}(2006)}]{diaz06}
{D{\'{\i}}az}, A.~J., \& {Roberts}, B. 2006,
  \href{http://dx.doi.org/10.1007/s11207-006-0137-y}{\JournalTitle{\solphys},
  236, 111}

\bibitem[{{Engvold}(2004)}]{engv04}
{Engvold}, O. 2004, \href{http://dx.doi.org/10.1017/S1743921304005575}{in IAU
  Symposium, Vol. 223, Multi-Wavelength Investigations of Solar Activity, ed.
  A.~V. {Stepanov}, E.~E. {Benevolenskaya}, \& A.~G. {Kosovichev}}, 187

\bibitem[{{Erd{\'e}lyi} \& {Fedun}(2007)}]{erde07}
{Erd{\'e}lyi}, R., \& {Fedun}, V. 2007,
  \href{http://dx.doi.org/10.1126/science.1153006}{\JournalTitle{Science}, 318,
  1572}

\bibitem[{{Jing} {et~al.}(2003){Jing}, {Lee}, {Spirock}, {Xu}, {Wang}, \&
  {Choe}}]{jing03}
{Jing}, J., {Lee}, J., {Spirock}, T.~J., {et~al.} 2003,
  \href{http://dx.doi.org/10.1086/373886}{\JournalTitle{\apjl}, 584, L103}

\bibitem[{{Jing} {et~al.}(2010){Jing}, {Yuan}, {Wiegelmann}, {Xu}, {Liu}, \&
  {Wang}}]{jing10}
{Jing}, J., {Yuan}, Y., {Wiegelmann}, T., {et~al.} 2010,
  \href{http://dx.doi.org/10.1088/2041-8205/719/1/L56}{\JournalTitle{\apjl},
  719, L56}

\bibitem[{{Joarder} \& {Roberts}(1992{\natexlab{a}})}]{joar92a}
{Joarder}, P.~S., \& {Roberts}, B. 1992{\natexlab{a}}, \JournalTitle{\aap},
  256, 264

\bibitem[{{Joarder} \& {Roberts}(1992{\natexlab{b}})}]{joar92b}
---. 1992{\natexlab{b}}, \JournalTitle{\aap}, 261, 625

\bibitem[{{Karpen} {et~al.}(2001){Karpen}, {Antiochos}, {Hohensee}, {Klimchuk},
  \& {MacNeice}}]{karp01}
{Karpen}, J.~T., {Antiochos}, S.~K., {Hohensee}, M., {Klimchuk}, J.~A., \&
  {MacNeice}, P.~J. 2001,
  \href{http://dx.doi.org/10.1086/320497}{\JournalTitle{\apjl}, 553, L85}

\bibitem[{{Karpen} {et~al.}(2006){Karpen}, {Antiochos}, \& {Klimchuk}}]{karp06}
{Karpen}, J.~T., {Antiochos}, S.~K., \& {Klimchuk}, J.~A. 2006,
  \href{http://dx.doi.org/10.1086/498237}{\JournalTitle{\apj}, 637, 531}

\bibitem[{{Karpen} {et~al.}(2005){Karpen}, {Tanner}, {Antiochos}, \&
  {DeVore}}]{karp05}
{Karpen}, J.~T., {Tanner}, S.~E.~M., {Antiochos}, S.~K., \& {DeVore}, C.~R.
  2005, \href{http://dx.doi.org/10.1086/497531}{\JournalTitle{\apj}, 635, 1319}

\bibitem[{{Keppens} {et~al.}(2012){Keppens}, {Meliani}, {van Marle}, {Delmont},
  {Vlasis}, \& {van der Holst}}]{kepp12}
{Keppens}, R., {Meliani}, Z., {van Marle}, A.~J., {et~al.} 2012,
  \href{http://dx.doi.org/10.1016/j.jcp.2011.01.020}{\JournalTitle{Journal of
  Computational Physics}, 231, 718}

\bibitem[{{Kippenhahn} \& {Schl{\"u}ter}(1957)}]{kipp57}
{Kippenhahn}, R., \& {Schl{\"u}ter}, A. 1957, \JournalTitle{\zap}, 43, 36

\bibitem[{{Li} \& {Zhang}(2012)}]{li12}
{Li}, T., \& {Zhang}, J. 2012,
  \href{http://dx.doi.org/10.1088/2041-8205/760/1/L10}{\JournalTitle{\apjl},
  760, L10}

\bibitem[{{Lin} {et~al.}(2007){Lin}, {Engvold}, {Rouppe van der Voort}, \& {van
  Noort}}]{lin07}
{Lin}, Y., {Engvold}, O., {Rouppe van der Voort}, L.~H.~M., \& {van Noort}, M.
  2007,
  \href{http://dx.doi.org/10.1007/s11207-007-0402-8}{\JournalTitle{\solphys},
  246, 65}

\bibitem[{{Liu} {et~al.}(2012){Liu}, {Ofman}, {Nitta}, {Aschwanden},
  {Schrijver}, {Title}, \& {Tarbell}}]{liu12}
{Liu}, W., {Ofman}, L., {Nitta}, N.~V., {et~al.} 2012,
  \href{http://dx.doi.org/10.1088/0004-637X/753/1/52}{\JournalTitle{\apj}, 753,
  52}

\bibitem[{{Luna} {et~al.}(2012){Luna}, {D{\'{\i}}az}, \& {Karpen}}]{luna12a}
{Luna}, M., {D{\'{\i}}az}, A.~J., \& {Karpen}, J. 2012,
  \href{http://dx.doi.org/10.1088/0004-637X/757/1/98}{\JournalTitle{\apj}, 757,
  98}

\bibitem[{{Luna} \& {Karpen}(2012)}]{luna12b}
{Luna}, M., \& {Karpen}, J. 2012,
  \href{http://dx.doi.org/10.1088/2041-8205/750/1/L1}{\JournalTitle{\apjl},
  750, L1}

\bibitem[{{Luna} {et~al.}(2016){Luna}, {Terradas}, {Khomenko}, {Collados}, \&
  {de Vicente}}]{luna16}
{Luna}, M., {Terradas}, J., {Khomenko}, E., {Collados}, M., \& {de Vicente}, A.
  2016,
  \href{http://dx.doi.org/10.3847/0004-637X/817/2/157}{\JournalTitle{\apj},
  817, 157}

\bibitem[{{Mackay} \& {van Ballegooijen}(2009)}]{mack09}
{Mackay}, D.~H., \& {van Ballegooijen}, A.~A. 2009,
  \href{http://dx.doi.org/10.1007/s11207-009-9468-9}{\JournalTitle{\solphys},
  260, 321}

\bibitem[{{Martens} \& {Zwaan}(2001)}]{mart01}
{Martens}, P.~C., \& {Zwaan}, C. 2001,
  \href{http://dx.doi.org/10.1086/322279}{\JournalTitle{\apj}, 558, 872}

\bibitem[{{Oliver} {et~al.}(2002){Oliver}, {Mann}, {Carballo}, {Franceschini},
  {Rowan-Robinson}, {Kontizas}, {Dapergolas}, {Kontizas}, {Verma}, {Elbaz},
  {Granato}, {Silva}, {Rigopoulou}, {Gonzalez-Serrano}, {Serjeant},
  {Efstathiou}, \& {van der Werf}}]{oliv02}
{Oliver}, S., {Mann}, R.~G., {Carballo}, R., {et~al.} 2002,
  \href{http://dx.doi.org/10.1046/j.1365-8711.2002.05309.x}{\JournalTitle{\mnras},
  332, 536}

\bibitem[{{Porth} {et~al.}(2014){Porth}, {Xia}, {Hendrix}, {Moschou}, \&
  {Keppens}}]{port14}
{Porth}, O., {Xia}, C., {Hendrix}, T., {Moschou}, S.~P., \& {Keppens}, R. 2014,
  \href{http://dx.doi.org/10.1088/0067-0049/214/1/4}{\JournalTitle{\apjs}, 214,
  4}

\bibitem[{{Ramsey} \& {Smith}(1966)}]{rams66}
{Ramsey}, H.~E., \& {Smith}, S.~F. 1966,
  \href{http://dx.doi.org/10.1086/109903}{\JournalTitle{\aj}, 71, 197}

\bibitem[{{Shen} {et~al.}(2014){Shen}, {Ichimoto}, {Ishii}, {Tian}, {Zhao}, \&
  {Shibata}}]{shen14}
{Shen}, Y., {Ichimoto}, K., {Ishii}, T.~T., {et~al.} 2014,
  \href{http://dx.doi.org/10.1088/0004-637X/786/2/151}{\JournalTitle{\apj},
  786, 151}

\bibitem[{{Soler} {et~al.}(2010){Soler}, {Arregui}, {Oliver}, \&
  {Ballester}}]{sole10}
{Soler}, R., {Arregui}, I., {Oliver}, R., \& {Ballester}, J.~L. 2010,
  \href{http://dx.doi.org/10.1088/0004-637X/722/2/1778}{\JournalTitle{\apj},
  722, 1778}

\bibitem[{{Tandberg-Hanssen}(1995)}]{tand95}
{Tandberg-Hanssen}, E., ed. 1995, Astrophysics and Space Science Library, Vol.
  199, {The nature of solar prominences}

\bibitem[{{Terradas} {et~al.}(2002){Terradas}, {Molowny-Horas}, {Wiehr},
  {Balthasar}, {Oliver}, \& {Ballester}}]{terr02}
{Terradas}, J., {Molowny-Horas}, R., {Wiehr}, E., {et~al.} 2002,
  \href{http://dx.doi.org/10.1051/0004-6361:20020967}{\JournalTitle{\aap}, 393,
  637}

\bibitem[{{Terradas} {et~al.}(2015){Terradas}, {Soler}, {Oliver}, \&
  {Ballester}}]{terr15}
{Terradas}, J., {Soler}, R., {Oliver}, R., \& {Ballester}, J.~L. 2015,
  \href{http://dx.doi.org/10.1088/2041-8205/802/2/L28}{\JournalTitle{\apjl},
  802, L28}

\bibitem[{{Vr{\v s}nak} {et~al.}(2007){Vr{\v s}nak}, {Veronig}, {Thalmann}, \&
  {{\v Z}ic}}]{vrsn07}
{Vr{\v s}nak}, B., {Veronig}, A.~M., {Thalmann}, J.~K., \& {{\v Z}ic}, T. 2007,
  \href{http://dx.doi.org/10.1051/0004-6361:20077668}{\JournalTitle{\aap}, 471,
  295}

\bibitem[{{Wang} \& {Sheeley}(1999)}]{wang99}
{Wang}, Y.-M., \& {Sheeley}, Jr., N.~R. 1999,
  \href{http://dx.doi.org/10.1086/311815}{\JournalTitle{\apjl}, 510, L157}

\bibitem[{{Withbroe}(1988)}]{with88}
{Withbroe}, G.~L. 1988,
  \href{http://dx.doi.org/10.1086/166015}{\JournalTitle{\apj}, 325, 442}

\bibitem[{{Withbroe} \& {Noyes}(1977)}]{with77}
{Withbroe}, G.~L., \& {Noyes}, R.~W. 1977,
  \href{http://dx.doi.org/10.1146/annurev.aa.15.090177.002051}{\JournalTitle{\araa},
  15, 363}

\bibitem[{{Xia} {et~al.}(2011){Xia}, {Chen}, {Keppens}, \& {van Marle}}]{xia11}
{Xia}, C., {Chen}, P.~F., {Keppens}, R., \& {van Marle}, A.~J. 2011,
  \href{http://dx.doi.org/10.1088/0004-637X/737/1/27}{\JournalTitle{\apj}, 737,
  27}

\bibitem[{{Yi} {et~al.}(1991){Yi}, {Engvold}, \& {Keil}}]{yi91}
{Yi}, Z., {Engvold}, O., \& {Keil}, S.~L. 1991,
  \href{http://dx.doi.org/10.1007/BF00159130}{\JournalTitle{\solphys}, 132, 63}

\bibitem[{{Zhang} {et~al.}(2012){Zhang}, {Chen}, {Xia}, \& {Keppens}}]{zhan12}
{Zhang}, Q.~M., {Chen}, P.~F., {Xia}, C., \& {Keppens}, R. 2012,
  \href{http://dx.doi.org/10.1051/0004-6361/201218786}{\JournalTitle{\aap},
  542, A52}

\bibitem[{{Zhang} {et~al.}(2013){Zhang}, {Chen}, {Xia}, {Keppens}, \&
  {Ji}}]{zhan13}
{Zhang}, Q.~M., {Chen}, P.~F., {Xia}, C., {Keppens}, R., \& {Ji}, H.~S. 2013,
  \href{http://dx.doi.org/10.1051/0004-6361/201220705}{\JournalTitle{\aap},
  554, A124}

\bibitem[{{Zhang} \& {Engvold}(1991)}]{zhan91}
{Zhang}, Y., \& {Engvold}, O. 1991,
  \href{http://dx.doi.org/10.1007/BF00152648}{\JournalTitle{\solphys}, 134,
  275}

\bibitem[{{Zhou} {et~al.}(2014){Zhou}, {Chen}, {Zhang}, \& {Fang}}]{zhou14}
{Zhou}, Y.-H., {Chen}, P.-F., {Zhang}, Q.-M., \& {Fang}, C. 2014,
  \href{http://dx.doi.org/10.1088/1674-4527/14/5/007}{\JournalTitle{Research in
  Astronomy and Astrophysics}, 14, 581}

\bibitem[{{Zou} {et~al.}(2017){Zou}, {Fang}, {Chen}, {Yang}, \& {Cao}}]{zou17}
{Zou}, P., {Fang}, C., {Chen}, P.~F., {Yang}, K., \& {Cao}, W. 2017,
  \href{http://dx.doi.org/10.3847/1538-4357/836/1/122}{\JournalTitle{\apj},
  836, 122}

\bibitem[{{Zou} {et~al.}(2016){Zou}, {Fang}, {Chen}, {Yang}, {Hao}, \&
  {Cao}}]{zou16}
{Zou}, P., {Fang}, C., {Chen}, P.~F., {et~al.} 2016,
  \href{http://dx.doi.org/10.3847/0004-637X/831/2/123}{\JournalTitle{\apj},
  831, 123}

\end{thebibliography}
\clearpage

\end{document}